\documentstyle[12pt]{article}
\begin{document}                                    
 \bibliographystyle{unsrt}

\title {Mass correction and gravitational energy radiation in black hole
perturbation theory}
\author{Reinaldo J. Gleiser}
\date{}
\maketitle

\begin{center} Facultad de Matem\'atica, Astronom\'{\i}a y F\'{\i}sica,
Universidad Nacional de C\'ordoba,\\ Ciudad
Universitaria, 5000 C\'ordoba, Argentina.
\end{center}

\begin{abstract}

Using second order black hole perturbation theory, we show that the
difference between the ADM mass and the final black hole mass, computed
to the lowest significant order, is equal, to the same order, to the
total gravitational radiation energy, obtained applying the Landau and
Lifschitz (pseudotensor) equation to the first order perturbation. This
result may be considered as a consistency check for the theory.

\end{abstract}
 
\vspace{1cm}

 
\vspace{1cm}

\section{Introduction}

The analysis of the motion of gravitating bodies in the strong field
region is, clearly, one of the most important problems in General
Relativity, since it is here where one would expect to find the most
pronounced effects, allowing, in principle, for a clear confrontation
of the theory with observation.  Perhaps the most important of these
effects, and the one to which the largest theoretical and experimental
efforts is being devoted nowadays, is the emission of gravitational
radiation through gravitational waves.  As is well known, however, and
in contrast with the situation in other theories,  e.g.
electromagnetism, the whole subject of gravitational waves and
gravitational radiation in the context of general relativity is a very
complex one, both from the point of view of their physical
interpretation as from that of the difficulties in constructing
solutions, (either exact or sufficiently approximate), appropriate for
the modeling of relevant astrophysical systems.

From very general arguments, one expects that the most prominent
sources of bursts of gravitational radiation, are the coalescences of
two astrophysical objects, leading to a single black hole as the final
result.  These same general ideas indicate that most of the radiation,
and therefore of the information to be observed at a large distance
from the region of emission, should come from the last stages of this
coalescence, where the system is close to, and rapidly approaching the
final black hole stage. This has lead, among other lines of attack, to
the idea that, given adequate initial data, one could consider these
last stages as corresponding to the evolution of the perturbations of a
suitable chosen black hole.

A successful application of this idea, attributed to Smarr, was carried
out by  Price and Pullin \cite{pullpri}, in the case of the head-on
collision of two black holes, in the close approximation, applying a
technique, originally developed by Regge and Wheeler \cite{R-W},and
Zerilli \cite{Zerilli}, (see also \cite{CPM}), devised precisely for
the analysis of linearized perturbations of black hole. By successful
we mean here that the results given perturbatively in \cite{pullpri},
are essentially identical to those obtained by ``exact'' numerical
solution of Einstein's equations, for the same initial data, for a
suitably chosen range of the parameters characterizing the problem.

It should be clear, however, that if the numerical results had not been
available, it would have been essentially impossible to asses the range
of validity of the perturbative results, not only because it relays
only on linearized theory, but also because of the intuitively
appealing, but non rigorous nature of the identification of what is
meant by the ``gravitational energy'' radiated by the system.

An important step in the direction of defining a range of validity for
the perturbative treatment was given in \cite{GNPP}, where the
formalism was extended to second order, with the idea that second order
corrections should serve as a sort of ``error bars'' on the first order
computations. This formalism was applied to the head-on collision of
black holes in \cite{prl}, showing a very remarkable agreement between
the ``exact'' result, the first order results and the associated
''error bars''.

A somewhat different application of black hole perturbation theory,
taken to second order, would be to analyze, instead of the
gravitational radiation, the change in the mass parameter of the black
hole. From a physical point of view, we expect that if an isolated
astrophysical system contains, at a given time, certain mass-energy,
and that if, after a long time, part of this mass-energy is radiated to
infinity, the sum of the mass-energy remaining in the system and that
radiated to infinity, should be equal to initial mass. In more rigorous
terms, the initial mass-energy, for an isolated system, is given by the
ADM mass. Since the process which we are envisioning is one where the
final result is a black hole, the final mass-energy is simply that of
the final black hole. The energy radiated to infinity should then be
equal to the difference between these two masses. This comparison would
provide an independent check of the validity of the computation of the
radiated energy, as given in, e.g., \cite{pullpri}.

In this note we consider the problem of computing the correction to the
mass parameter in the lowest non trivial order in black hole
perturbation theory. After reviewing briefly the formalism, we derive a
general equation for  the relevant metric perturbation function, and,
under some general restrictions, and using the Zerilli equation, obtain
the expression for the mass correction, in terms of the Zerilli
function.  It is remarkable that the resulting expression is identical
to that obtained from the Landau-Lifschitz pseudotensor for the total
gravitational energy radiation, as given in \cite{CPM}, although this
is not mentioned anywhere in the derivations. In some sense, we might
even consider this reassuring proof of consistency of the perturbation
expansion, as a ``derivation'' of the radiation formula, since, as will
be seen below, only general properties of the solutions of Einstein's
equations are used in the proofs, without any reference to
asymptotically flat or radiation gauges. We should, however, make clear
that it is not the intention of this article to delve into the
difficult problem of the definition of the energy in general
relativity, and of the relative merits of the different
``pseudotensors'' or ``complexes'' that have been discussed in the
literature \cite{aguirregabiria}. In particular, the comparison is made
only to the expression for the integrated power, and it is known that
different prescriptions may give here the same final answer. We also
remark that the equivalence between the Bondi energy flux and that
given by the Landau-Lifschitz complex, in suitable asymptotically flat
space-times, has been shown in \cite{persides}.

\section{The perturbative expansion}

The essence of the perturbation method is the expansion of the metric
in the form 
\begin{equation} 
g_{\mu \nu}(\epsilon) = g_{\mu \nu}^{(0)}
+\epsilon g_{\mu \nu}^{(1)} +\epsilon^2 g_{\mu \nu}^{(2)} +
{\cal{O}}(\epsilon^3) 
\end{equation} 
where $g_{\mu \nu}(\epsilon)$ represents a one parameter family of
solutions, and $g_{\mu \nu}^{(0)}$ some known exact solution of
Einstein's equations, leading to an expansion of the Einstein equation
into a hierarchy of equations, ordered also by the parameter
$\epsilon$, where the solution of each order, considered as an initial
value problem, requires the solution of all previous ones
\cite{GNPP}. Some relevant issues, including gauge invariance, are
discussed in more detail in \cite{gau}. (See also \cite{bruni}.)

In our case, $g_{\mu \nu}^{(0)}$ corresponds to the Schwarzschild black
hole, and $\epsilon$ to some parameter characterizing the departure of
the initial data from that of a black hole. A particularly suitable
framework for this problem is given by the Regge-Wheeler \cite{R-W}
formalism, where the perturbations to any given order are given as a
multipolar expansion in the angular variables $\theta$, and $\phi$. The
multipoles are, in turn, separated into even and odd type, for any
given $L$, the order of the multipole. The $L=0$, (monopole), terms
contain information on the mass of the system. We shall assume that the
leading (first order in $\epsilon$) part of the perturbation
corresponds to the $L=2$, even, (quadrupole) terms. (This is, for
instance, the case considered in \cite{pullpri}). Therefore, all other
terms are, at least, of order $\epsilon^2$
 
We choose from the outset a Regge - Wheeler gauge \cite{R-W}. This
means that the non vanishing metric metric coefficients are
\begin{eqnarray}
g_{tt} & = &-\left(1-\frac{2m}{r}\right)\left[1-\epsilon H_0^{(1)}(t,r)
P_2(\theta)-\epsilon^2 H_0^{(2)}P_0\right]\nonumber \\
g_{tr} & = &\epsilon H_1^{(1)}(t,r) P_2(\theta) \nonumber \\
g_{rr} & = &\left(1-\frac{2m}{r}\right)^{-1}\left[1+\epsilon
H_2^{(1)}(t,r)
P_2(\theta)+\epsilon^2 H_2^{(2)}P_0\right]\nonumber \\
\label{metrica}
g_{\theta \theta} & = &r^2[1+\epsilon K^{(1)}(t,r) P_2(\theta)] \\
g_{\phi \phi} & = &r^2\sin(\theta)^2[1+\epsilon K^{(1)}(t,r)
P_2(\theta)]  \nonumber
\end{eqnarray}
where  the $P_L$, $L=0,2$ are Legendre polynomial in $\cos(\theta)$.
The factor $\epsilon$ makes explicit the perturbation order. The
functions $H_0^{(1)}(t,r)$, $H_1^{(1)}(t,r)$, $H_2^{(1)}(t,r)$, and
$K^{(1)}(t,r)$ characterize the first order, $L=2$, even,
perturbations, while, $H_0^{(2)}(t,r)$, and $H_2^{(2)}(t,r)$ describe
the $L=0$, second order perturbations. The general second order
expansion contains terms also for $L=2$ and $L=4$, but these are
decoupled, to this order, from each other, and from the $L=0$ terms,
and we shall not consider them here.

We remark that although, in general, the $L=0$ perturbations would also
contain  terms of the form
\begin{eqnarray}
g_{tr} & = &\epsilon^2 H_1^{(2)}(t,r) P_0(\theta) \nonumber \\
g_{\theta \theta} & = &r^2\epsilon^2 K^{(2)}(t,r) P_0(\theta) \\
g_{\phi \phi} & = &r^2\sin(\theta)^2\epsilon^2 K^{(2)}(t,r)
P_0(\theta)  \nonumber
\end{eqnarray}
it is always possible to choose a gauge where $H_1^{(2)}(t,r)=0$, and $
K^{(2)}(t,r)=0$, so we make that simplifying choice. Furthermore, with
this choice, if we assume that the first order perturbations vanish,
the solution of Einstein's equations for $H_2^{(2)}$ is  time
independent, and  of the form
\begin{equation}
H_2^{(2)} = {C \over r-2M}
\end{equation}
where $C$ is a constant.

Now, if we consider as the only "perturbation" a change in mass $\delta
M$, it can be seen that
\begin{equation}
H_2^{(2)} = {2 \delta M \over r-2M}
\end{equation}
Therefore, we identify the constant $C$ with twice the correction to
the mass.

Taking into account the different powers in $\epsilon$, and in a manner
analogous to that in \cite{GNPP}, the Einstein equations for
(\ref{metrica}) separate into a linear set of equations for
$H_0^{(1)}(t,r)$, $H_1^{(1)}(t,r)$, $H_2^{(1)}(t,r)$, and
$K^{(1)}(t,r)$, corresponding to terms linear in $\epsilon$, and a set
of equations linear in $H_0^{(2)}(t,r)$, and $H_2^{(2)}(t,r)$, but
containing ``source''-like terms, quadratic in the first order
functions.

Regarding the first order functions, we recall that the general
solution to the first order equations can be given in terms of the
Zerilli function. Namely, any set $H_0^{(1)}(t,r)$, $H_1^{(1)}(t,r)$,
$H_2^{(1)}(t,r)$, $K^{(1)}(t,r)$ of solutions of these equations can be
written in the form
\begin{eqnarray}
\label{K1}
K^{(1)}(t,r) & = & 6 {r^2+rM +M^2 \over
r^2(2r+3M)}\psi^{(1)}(t,r)+\left(1-2 {M
\over r} \right) {\partial \psi^{(1)}(t,r)\over \partial r}   \\
\label{H2}
H^{(1)}_2(t,r) & = & {\partial \over \partial r}\left[{2r^2-6rM-3M^2
\over
r(2r+3M)} \psi^{(1)}(t,r)+(r-2M) {\partial \psi^{(1)}(t,r) \over
\partial r}
\right] - K^{(1)}(t,r)   \\
\label{H1}
H^{(1)}_1(t,r) & = & {2r^2-6rM-3M^2 \over (r-2M)(2r+3M)} {\partial
\psi^{(1)}(t,r)
\over \partial t}+r {\partial^2 \psi^{(1)}(t,r) \over \partial r
\partial t}
 \\
H^{(1)}_0(t,r) & = & H^{(1)}_2(t,r) 
\end{eqnarray}
where
$\psi^{(1)}(t,r)$ is a solution of the Zerilli equation:
\begin{equation}
{\partial^2 \psi^{(1)}(t,r) \over \partial {r^*}^2} - {\partial^2
\psi^{(1)}(t,r)
\over \partial t^2} -V(r^*) \psi^{(1)}(t,r) =0
\label{Zer21}
\end{equation}
where
\begin{equation}
r^*=r+2M \ln[r/(2M)-1]
\end{equation}
and
\begin{equation}
V(r) = 6\left(1-2{M \over r}\right){4 r^3+4 r^2 M+6 r M^2+3 M^3
\over r^3 (2 r+3 M)^2 }
\end{equation}

We remark that, for sufficiently large $r$, (\ref{Zer21}) approaches
the wave equation, and, therefore, for sufficiently small initial data,
$\psi^{(1)}(t,r)$ approaches a function of $t-r*$, for large $r$ and
$t$.

Going now to the second order in $\epsilon$, $L=0$, perturbations, we
find that there are four non trivial equations (written as $R_{\mu \nu}
=0$, where $R_{\mu \nu}$ is the Ricci tensor), for the second order
functions. The equation $R_{tr} =0$ takes the form
\begin{equation}
{\partial H_2^{(2)}(t,r) \over \partial t} = {\cal{S}}
\end{equation}
where ${\cal{S}}$ is quadratic in the first order perturbations. We
assume a choice of initial data, (for $t=0$), such that $M$ is equal to
the ADM mass, and, therefore, we should have
\begin{equation}
\lim_{r \rightarrow \infty} \left[ r H_2^{(2)}(0,r) \right] = 0
\end{equation}
so that there is no correction to the mass for $t=0$.

Since on the other hand, for large $t$, the solution should approach
the static black hole configuration, we should have
\begin{equation}
\lim_{t \rightarrow \infty} \left[  H_2^{(2)}(t,r) \right] = {2 \delta
M \over r-2M}
\end{equation}
and we find that
\begin{equation}
\int_0^{\infty}{\partial H_2^{(2)}(t,r) \over \partial t} dt = {2
\delta M \over r-2M} - H_2^{(2)}(0,r)
\end{equation}
and, then,
\begin{equation}
\delta M  = {1 \over 2} \lim_{r \rightarrow \infty} \left[ r
\int_0^{\infty} {\cal{S}} dt \right]
\end{equation}

Using the asymptotic properties of the Zerilli function, and a fair
amount of algebra, (some details are given in the Appendix), we find,
\begin{equation}
\delta M  = - \lim_{r \rightarrow \infty} \left[{3 \over 10}   \int_0^{\infty} \left| {\partial \psi \over \partial t} \right|^2 dt \right]
\label{deltaM} 
\end{equation}
which expresses the change in mass in terms of the first order
perturbations, through the Zerilli function. But the right hand side of
this equation is also immediately recognized as precisely (minus) the
total gravitational radiation energy, computed from the Landau -
Lifschitz (pseudotensor) equation \cite{CPM}. 

\section{Comments}

The main result of the present analysis, indicated by Eq.
(\ref{deltaM}) is interesting in several respects. To begin with, we
remark that it was obtained in the framework of {\em second order
perturbation theory}, namely, the presence of the ``source terms'',
containing the contributions from the lower (first) order perturbations
was crucial in its derivation. Second, we find an expression for the
radiated energy, working only in the Regge-Wheeler gauge, that is
totally independent of any consideration of asymptotically flat or
radiation  gauges and, or identifications of gravitational wave
amplitudes. Furthermore, the expression, although computed in second
order perturbation theory, contains only information from the first
order quantities. Thus, in a sense, it may be considered as a
``derivation'' of the radiation formula to be used in first order
perturbation theory. That this equation coincides with the
Landau-Lifschitz pseudotensor prescription, may then be taken as a
reassuring (although partial) proof of the physical consistency of the
black hole perturbation treatment.

Finally we remark that the proof in this note has been limited to a
restricted, although relevant, set of perturbations, since only the
first order, $L=2$ even case was considered. This choice was made
mainly for simplicity and we expect that an analogous result
should hold in the general case. This will be considered elsewhere.

\section*{Appendix}

We include in this Appendix some computational details. First we notice
that, from $R_{tr} =0$, after some simplifications using the first
order equations,  we find
\begin{eqnarray}
r H_2^{(2)},_t & = &
{r^2\over 10}   \left( H_1^{(1)}  ,_r    H_1^{(1)},_t
-     H_2^{(1)}  ,_r    H_2^{(1)}  ,_t \right)
-{r\over 5} \left(  K^{(1)}   H_2^{(1)}  ,_t + H_1^{(1)}
H_1^{(1)},_t-  H_2^{(1)}      H_2^{(1)}  ,_t \right)  \nonumber \\
\label{Ap1}
& &  
-{2\over 5}  K^{(1)}    H_1^{(1)}  -{r-2M\over 5r}  H_1^{(1)}
H_2^{(1)}
+{M\over 5} \left(H_1^{(1)} H_2^{(1)},_r -  H_1^{(1)},_r H_2^{(1)}
\right)\;.
\end{eqnarray}

Assuming an outgoing wave boundary condition for large $r$,
(appropriate for perturbative initial data), one can show that
$\psi(t,r)$ admits the following asymptotic expansion
\begin{equation}
\psi(t,r) = {1 \over 3} F^{\{3\}} +{1\over r} F^{\{1\}} +{1 \over r^2}
\left( F-M F^{\{1\}} \right) + {1 \over r^3} \left( {7 M^2 \over 4}
F^{\{1\}} - MF \right) + {\cal{O}}(1/r^4)
\label{asympsi}
\end{equation}
where ${\cal{O}}(1/r^4)$ means an expression whose absolute value is
bounded by $A/r^4$, with $A$ some positive constant,  $F=F(t-r^*)$, is
an arbitrary real function, and $F^{\{n\}}(x) = d^n F / dx^n$. We
assume also that $F(x)$, $F^{\{1\}}(x)$, and $F^{\{2\}}(x)$ are
bounded, that  $F^{\{3\}}(x)$, is square integrable, and that and
$$\lim_{x \rightarrow +\infty} x^2 F^{\{3\}}(-x)F^{\{4\}}(-x) = 0$$
(These are essentially a ``finite energy'', plus ``smoothness at
infinity'' conditions).
 
Replacing (\ref{K1}), (\ref{H2}), and (\ref{H1}) in (\ref{Ap1}), and
expanding the resulting expression after replacing the asymptotic
expression (\ref{asympsi}) for $\psi$, we find
\begin{eqnarray}
r {\partial H_2^{(2)} \over \partial t}  & = &  -{1 \over 15}(
F^{\{3\}})^2+ {r^2 \over 45}  {\partial \over \partial t} \left(
F^{\{3\}} F^{\{4\}}\right)
+ {4r \over 45}{\partial \over \partial t} \left[ {M \over 2} F^{\{3\}}
F^{\{4\}} + ( F^{\{3\}})^2 \right] \nonumber \\
\label{ap2}
& & + {\partial \over \partial t} \left[ {4M^2 \over 45} F^{\{3\}}
F^{\{4\}}
+{5M \over 18} ( F^{\{3\}})^2 + {2 \over 15} F^{\{2\}} F^{\{3\}}
\right] + {\cal{O}}(1/r)
\end{eqnarray}

From the assumptions made above on $F$, it is clear that the time
integrals of the second, third and fourth terms on the right hand side
of (\ref{ap2}), required to obtain $H_2^{(2)}$ vanish in the limit of
large $r$. We therefore have
\begin{equation}
 \lim_{r \rightarrow \infty} \left[r  \int_0^{\infty} {\partial
 H_2^{(2)} \over \partial t} dt \right]
= - {1 \over 15}     \int_0^{\infty} ( F^{\{3\}})^2  dt 
\end{equation}
which leads immediately to (\ref{deltaM}).

\section*{Acknowledgements}

I wish to thank R. H. Price, J. Pullin, and C. O. Nicasio for their
helpful comments and suggestions. This work was supported in part by
the National University of C\'ordoba, by grant NSF-INT-9512894, and
grants from CONICET and CONICOR (Argentina). The author is a member of
CONICET.

\end{document}